\documentstyle[12pt,aaspp4,flushrt]{article}
\begin{document}
\title{ 
The R-Mode Oscillations in Relativistic Rotating Stars 
}
%
\author{
    Yasufumi Kojima and  Masayasu Hosonuma
}
\affil{ 
Department of Physics, Hiroshima University,  
         Higashi-Hiroshima 739-8526, Japan  
           }
\authoremail{kojima@theo.phys.sci.hiroshima-u.ac.jp}
%
\begin{abstract}
   The axial mode oscillations are examined for
relativistic rotating stars with uniform angular velocity.
Using the slow rotation formalism and the Cowling 
approximation, we have derived the equations governing  the 
r-mode oscillations
up to the second order with respect to the rotation.
In the lowest order, the allowed range of the
frequencies  is determined, but corresponding spatial 
function is arbitrary. 
The spatial function can be decomposed
in non-barotropic region
by a set of functions associated with the differential equation 
of the second-order corrections.
The  equation  however becomes singular in
barotropic region, and a single function can be selected 
to describe the spatial perturbation of the lowest order.
The frame dragging effect among the relativistic effects
may be significant, as it results in rather broad spectrum 
of the r-mode frequency 
unlike in the Newtonian first-order calculation.
\end{abstract}


\section{Introduction}

In recent years, the r-mode oscillations in rotating
stars have been found to be significant implications.
The axial oscillations are unstable by the 
gravitational radiation reaction (\cite{an98}, \cite{frmo98}). 
The mechanism of the instability can be understood by the 
generic argument for the gravitational radiation driven 
instability, so-called CFS-instability, 
which is originally examined for the polar perturbations, 
(\cite{ch70}, \cite{frsc78}, \cite{fr78}). 
All rotating stars become unstable in the absence of
internal fluid dissipation, irrespective of the parity modes.
Viscosity however damps out the oscillations, and stabilizes them 
in general.
The polar f-mode instability is believed to act 
only in rapidly rotating neutron stars (\cite{li95}). 
The axial instability is however found to set in even 
in much more slowly rotating cases, 
and to play an important role on the evolution of 
hot newly-born neutron stars (\cite{liow98}, \cite{ankosc98}).
The instability  carries away most of angular momentum and 
rotational energy of the stars by the gravitational radiation.
The gravitational wave emitted during the spin-down process is
expected to be one of the promising sources for the laser 
interferometer  gravitational wave detectors (\cite{owetal98}).

Most of the estimates for the r-mode instability 
are based on the Newtonian calculations.
That is, the oscillation frequencies are determined by 
inviscid hydrodynamics under the Newtonian gravity, and 
the gravitational radiation reaction is incorporated by
evaluating the (current) multipole moments.
Relativity has great influence on stellar structures,
redshift in oscillation frequencies, frame dragging,
radiations and so on.
It is important to explore the relativistic effects,
which may not change the general features of the 
oscillations, but  shift the critical angular velocity.
The relativistic calculation is not so easy task, even using 
the linear perturbation method concerning the oscillation 
amplitude.
Both relativity and rotation 
complicate the problems considerably.

As the first step toward a clear understanding of
relativistic corrections, 
we have examined  the r-mode oscillations
with two approximations, the slow rotation approximation 
(\cite{ha67})
and the Cowling approximation (\cite{co41}).
The rotation is assumed to be slow, and treated as small 
perturbation from the non-rotating case.
We also assume that the perturbation of the 
gravity can be neglected in the oscillations.
The accuracy of the Cowling approximation are tested
in the rotating relativistic system
(\cite{fi88}, \cite{lisp90}, \cite{yoko97}).
The calculations give the same qualitative results and
reasonable accuracy of the oscillation frequencies,
as in the Newtonian stellar pulsation theory
(e.g., \cite{co80}).
    In Section 2, we summarize the basic equations 
for the inviscid fluid in slowly rotating  relativistic stars
and the perturbation scheme to solve them.
The r-mode solution is given at the lowest order with 
respect to the rotational parameter in Section 3.
The mode can not be specified by a single frequency unlike 
in the Newtonian r-mode oscillation,
although the frequency is bounded to a certain range.
The spatial function of the mode is arbitrary at this order. 
In order to determine the radial structure, we need the 
rotational  corrections up to the third order  
in the background.
In Section 4, we include the rotational effects, 
and derive the equation
governing the  r-mode oscillations in non-barotropic region.
The resultant equation becomes singular in the barotropic 
case, since the second-order differential term vanishes.
In Section 5, we separately consider for the barotropic stars 
the mode function and the correction to the leading order. 
In Section 6, we discuss the implication of our results.
We use the geometrical units of $c=G=1. $
%

\section{Pulsation equations of axial mode}

We consider a slowly rotating star with a uniform angular 
velocity $ \Omega \sim O( \varepsilon ) $.
The rotational effects can be treated as the corrections to the 
non-rotating spherical star.
We will  take account of the corrections up to the third order in 
$\varepsilon$.
The metric tensor for describing the 
stationary axisymmetric  star is given by
(\cite{ha67},\cite{chmi74}, \cite{qu76})
\begin{eqnarray}
 ds^2
  &=&
  -e^{\nu}\left[1+2(h_0 + h_2P_2)\right]dt^2
  +e^{\lambda}\left[1+\frac{2e^{\lambda}}{r}
     \left( m_0 + m_2P_2 \right) \right]dr^2
\nonumber
\\
&&~
  +r^2(1 + 2k_2P_2)
   \left\{
    d\theta^2
    +
    \sin^2 \theta \left[ d\phi
             -
             \left(
              \omega + W_1 - W_3 \frac{1}{\sin \theta} \frac{d P_{3}}{d \theta} 
              \right)dt
             \right]^2
    \right\},
\label{eq.metr}
\end{eqnarray}
where $P_l(\cos \theta )  ~(l=2,3)$
denotes  the Legendre's polynomial of degree $l$.
The metric functions except $ g_{t \phi}$
should be expanded by an even power of $\varepsilon$,
while $ g_{t \phi}$ by an odd power of $\varepsilon$
due to the rotational symmetry, i.e., 
$ \omega \sim O( \varepsilon ),$ 
$(h_0, h_2, m_0, m_2, k_2 ) \sim O( \varepsilon^2 )$
and $ (W_1, W_3) \sim O( \varepsilon^3 ).$ 
These quantities are functions of the radial coordinate $r$.

The equilibrium state is assumed to be described 
with perfect-fluid stress-energy tensor.
The 4-velocity of the fluid element inside the star 
has the components, which are correct to $O( \varepsilon^3 ),$ 
\begin{equation}
 U^{\phi}   =  \Omega U^{t} ,
~~~~
 U^{t}  =
 ( - g_{tt} - 2\Omega g_{t\phi} -\Omega^2 g_{\phi\phi})^{-1/2} .
\end{equation}
The pressure and density distributions
are subject to the centrifugal deformation,
which is the effect of $ O(\varepsilon ^2 ). $
These distributions are expressed as 
\begin{equation} 
 p = p_0 (r) + \{ p_{20}(r) + p_{22}(r) P_2 (\cos \theta) \},
\end{equation}
\begin{equation}
 \rho = \rho_0 (r) +
 \{ \rho_{20} (r) + \rho_{22} (r) P_2 (\cos \theta) \},
\label{eq.dens}
\end{equation}
where $p_0 $ and $\rho_0$ are the
values for the non-rotating star, and the quantities in the braces
are the rotational corrections.
The non-rotating spherical configuration and
the rotational corrections in eqs.(\ref{eq.metr})-(\ref{eq.dens}) 
are determined by successively solving the perturbed Einstein 
equations with the same power of $ \varepsilon $
(\cite{ha67},\cite{chmi74}, \cite{qu76}).

  We consider the linear perturbations of the equilibrium fluid state.
We use the Eulerian change of the pressure,  
density and 4-velocity, which are represented by 
$\delta p$, $ \delta \rho$ and  $ \delta U_\nu$.
The perturbations of the gravitational field are neglected in 
the Cowling approximation.
It is convenient in the following calculation to expand these 
functions in terms of the spherical harmonics $ Y_{lm} ,$
\begin{eqnarray}
\delta p  &=&
\sum_{lm} \delta p_{lm}(t,r) Y_{lm}( \theta , \phi) ,
\\
\delta \rho
  &=&
 \sum_{lm}\delta \rho_{lm} (t,r) Y_{lm} ( \theta , \phi) ,
\\
 (\rho + p)\delta U_{r}
  &=&
  e^{\nu/2}\sum_{lm} R_{lm} (t,r) Y_{lm}( \theta , \phi) ,
\\
 (\rho + p)\delta U_{\theta}
  &=&
  e^{\nu/2}\sum_{lm}
  \left[
   V_{lm} (t,r) \partial_{\theta} Y_{lm}( \theta , \phi)
   -
   \frac{U_{lm}(t,r)}{ \sin \theta }
   \partial_{\phi} Y_{lm} ( \theta , \phi)
   \right],
\\
 (\rho + p)\delta U_{\phi}
  &=&
  e^{\nu/2}\sum_{lm}
  \left[
   V_{lm} (t,r)\partial_{\phi} Y_{lm} ( \theta , \phi)
   +
   U_{lm} (t,r) \sin \theta \partial_{\theta}Y_{lm} ( \theta , \phi)
   \right],
\\
 \delta U_{t}
  &=&
  -  \Omega \delta U_{\phi} .
\end{eqnarray}
With these definitions, the pulsation equations are derived from 
the conservation laws of the perturbed energy-momentum,  
\begin{equation}
  \delta T_{\mu;\nu}^{\nu} = 0, 
\label{eq.cons}
\end{equation}
where
\begin{equation}
  \delta T_\mu ^\nu    =
  ( \delta \rho + \delta p )  U_{\mu} U^{\nu}
 + \left[
   (\rho + p) \delta U_{\mu} U^{\nu}
   +
   (\rho + p) \delta U^{\nu} U_{\mu}
   \right]
 +  \delta p \delta_\mu ^\nu .
\end{equation}
Since we assume that the perturbation is adiabatic, 
the thermodynamic relation between
the perturbed pressure and density can be written as
\begin{equation}
 \delta p  + \xi \cdot \nabla p 
= \frac{\Gamma p }{p + \rho }
  \left(  \delta \rho + \xi \cdot \nabla  \rho   \right ) ,
\label{therm0}
\end{equation}
where  $ \Gamma $ is the adiabatic index and
$ \xi $ is the Lagrange displacement. 

We will solve eqs.(\ref{eq.cons}) and (\ref{therm0})
by the expansion of $\varepsilon$.
In a spherical star, the perturbations decouple into
purely polar and purely axial modes for each $ l$ and $m$. 
A set of the functions  ${\cal P}_{lm}  \equiv $
$ ( \delta p_{lm},  \delta \rho_{lm}, R_{lm} ,  V_{lm} )$ 
describes the polar mode, while  the function 
${\cal A}_{lm} \equiv $ $U_{lm} $
describes the axial mode.
In the presence of the rotation, 
the mode will be mixed among the terms with different  $l$, 
while the mode can still be specified by a single $m$
(e.g., \cite{ko92}, \cite{ko97}).
The coupled equations are schematically expressed as
\begin{equation}
 0 =  [ {\cal A}_{lm} ] +  
  {\cal E} \times  [ {\cal P}_{l\pm 1m} ]+
  {\cal E}^2 \times [ {\cal A}_{lm} ,  {\cal A} _{l\pm 2m} ]
  + \cdots,
\label{eq.axex}
\end{equation}
\begin{equation}
 0 =  [ {\cal P}_{lm} ]+ 
  {\cal E} \times [ {\cal A}_{l\pm 1m} ]+
  {\cal E}^2 \times [ {\cal P}_{lm} ,  {\cal P} _{l\pm 2m} ]
  + \cdots,
\label{eq.plex}
\end{equation}
where the symbol ${\cal E} $ means some functions of order 
$ \varepsilon ,$  and
the square bracket formally represents the relation 
among the axial perturbation function ${\cal A}_{lm} $,
or the polar perturbation function ${\cal P}_{lm}  $
appeared therein.
We also assume that 
the time variation of the oscillation is slow and
proportional to $\Omega, $ i.e., 
$\partial _t \sim \Omega \sim  O(\varepsilon) .$
This is true in the r-mode oscillation, as will be 
confirmed soon.
We look for the mode which is described by a single
axial function with index $(l,m)$ in the
limit of $\varepsilon \to 0.$
That is,  the polar part should vanish, while
the axial part becomes finite. Hence,  the
perturbation functions are expanded as 
\begin{equation}
    {\cal A}_{lm} =  {\cal A}_{lm} ^{(1)}+  
 \varepsilon ^2 {\cal A}_{lm} ^{(2)}  + \cdots,
~~~~
    {\cal P}_{lm} =  \varepsilon ( 
{\cal P}_{lm} ^{(1)}+  \varepsilon ^2 {\cal P}_{lm} ^{(2)} + \cdots ) .
\end{equation}
Substituting these functions into 
eqs.(\ref{eq.axex}) and (\ref{eq.plex}),
and comparing the coefficients of $ \varepsilon^n (n=0,1,2)$, 
we have 
\begin{eqnarray}
0 &=&  [ {\cal A}_{lm} ^{(1)} ],
\label{eq.lwst}
\\
0 &=&  [ \varepsilon {\cal P}_{l\pm1m} ^{(1)}  +
{\cal E} \times  {\cal A}_{lm} ^{(1)} ],
\label{eq.pol}
\\
0 &=&   [  \varepsilon ^2 {\cal A}_{lm} ^{(2)} ]    + 
{\cal E} \times [ \varepsilon  {\cal P}_{l\pm 1m} ^{(1)}]
+ {\cal E}^2 \times 
[ {\cal A}_{l m} ^{(1)},  {\cal A}_{l\pm 2m} ^{(1)} ]
= [ \varepsilon ^2 {\cal A}_{lm} ^{(2)}  +
 {\cal E}^2 \times  {\cal A}_{l m} ^{(1)}] ,
\label{eq.2nd}
\end{eqnarray}
where we have used  ${\cal A}_{l\pm 2m} ^{(1)} =0$ and
the relation (\ref{eq.pol}) in the last part of (\ref{eq.2nd}). 
Equation (\ref{eq.lwst}) represents the axial oscillation at the
lowest order, which can be specified by  $ U_{lm} ^{(1)} .$
Equation (\ref{eq.2nd}) is the second-order form of it, and
the term $ {\cal E}^2 \times {\cal A}_{l m} ^{(1)} $
can be regarded as the rotational corrections. 
We will show the explicit forms corresponding to
eqs.(\ref{eq.lwst})-(\ref{eq.2nd}) in subsequent sections.
%

\section{First-order solution}

The leading term of eq.(\ref{eq.cons}) is reduced to
\begin{equation}
  \dot{U}_{lm} ^{(1)} - im \chi U_{lm} ^{(1)} = 0 ,
\label{lap}
\end{equation}
where 
\begin{equation}
 \chi = \frac{2}{l(l+1)} \varpi 
      = \frac{2}{l(l+1)} (\Omega -\omega) , 
\end{equation}
and a dot denotes time derivative in the co-rotating frame, 
i.e., $\dot{U}_{lm} = (\partial_t + im \Omega) U_{lm} $.
The evolution of the perturbation can be 
solved by the Laplace transformation,
\begin{equation}
 u(s,r) = \int _0 ^\infty U_{lm} ^{(1)} (t,r) e^{-st} dt .
\end{equation}
The Laplace transformation of eq.(\ref{lap})   is written as
\begin{equation}
  \left( s + im (\Omega - \chi) \right ) u(s,r) - f_{lm} ^{(1)} (r)=0,
\end{equation}
where $f_{lm} ^{(1)}$ describes the initial disturbance 
at $t=0$.
After  solving  $ u$ and using the inverse transformation,
the solution in $t$-domain  is easily constructed as
\begin{equation}
  U_{lm} ^{(1)} (t,r) = \int  
f_{lm} ^{(1)} (r)  \frac{ e^{st} }{ s + im (\Omega - \chi) } ds = 
f_{lm} ^{(1)} (r) e^{ -im (\Omega-\chi) t} H(t),
\label{eq.1st}
\end{equation}
where $ H(t) $ is the Heaviside step function. 
We will consider $ t > 0 $ region only, so that 
the function $ H(t) $ may well be omitted from now on.  
For the Newtonian star, 
$m( \Omega - \chi) $  becomes a constant 
\begin{equation}
 \sigma _N =\left( 1- \frac{2}{l(l+1)} \right ) m \Omega .
\label{eq.nwfr}
\end{equation}
This is the r-mode frequency measured in the non-rotating frame
(\cite{papa78}, \cite{prbero81}, \cite{sa82}).
In the relativistic stars, 
$\varpi $  is monotonically increasing function of 
$r $, $\varpi _0 \leq  \varpi \leq \varpi_R .$ 
The possible range of the r-mode frequency is spread out.
If one regards eq.(\ref{eq.1st}) as the sum of the Fourier 
component $e^{- i \sigma t},$
then the spectrum is continuous in the range
\begin{equation} 
  \left( 1- \frac{2}{l(l+1)} \frac{\varpi _R}{\Omega} 
   \right )m \Omega  
   \leq  \sigma   \leq
   \left( 1- \frac{2}{l(l+1)} \frac{\varpi _0}{\Omega} 
    \right ) m \Omega  .
\label{eq.rng}
\end{equation}
This result is the same\footnote{
Note that this conclusion is derived within 
the lowest-order approximation of the perturbation scheme.
The consistency in the higher order will affect it for 
barotropic case as will be shown in the following sections.},
even if metric perturbations are considered (\cite{ko98}).
The time-dependence is determined, whereas the radial 
dependence $ f_{lm} ^{(1)}  $ is arbitrary at this order. 
The function $f_{lm} ^{(1)}  $ in eq.(\ref{eq.1st})  
is constrained by the equation of motions for the polar part,
as will be shown in the subsequent sections.
In this manner,
the perturbation scheme (\ref{eq.lwst})-(\ref{eq.2nd})
is degenerate perturbation.

\section{Second-order equation}

The pressure and density perturbations are induced by the 
rotation, while the axial mode in the non-rotating stars
is never coupled to them.
We may assume that the perturbation can be specified by
a single function $  U_{lm} ^{(1)}$ at the leading order,
since general case can be described by the  linear combination.
The axial part induces the polar parts with index $(l\pm 1, m)$
according to the perturbation scheme (\ref{eq.pol}). 
%
%
It is straightforward to solve $\delta p_{l\pm1 m},$
and  $\delta \rho_{l\pm1 m}$  from  two components of 
eq.(\ref{eq.cons}) in terms of the first-order corrections
and $  U_{lm} ^{(1)} $.
The explicit  results are 
\begin{eqnarray}
 \delta p_{l\pm1 m} 
   &=& 
   2 S_{\pm}\varpi U_{lm} ^{(1)},
\label{eq.pr}
\\
 \delta\rho_{l\pm1 m}
  &=&
  -4 S_{\pm} 
   \frac{ e^{-\nu/2} }{\nu'}
   ( e^{\nu/2} \varpi U_{lm}  ^{(1)})'
   +2 T_{\pm} 
    \frac{ e^{\nu} }{r^2 \nu'} 
    ( r^2 \varpi e^{-\nu} )' U_{lm} ^{(1)} ,
\label{eq.dn}
\end{eqnarray}
where
\begin{eqnarray}
  S_{+}  &=&   \frac{l}{l + 1} Q_{+} ,
~~~~
  S_{-} =   \frac{l + 1}{l} Q_{-} ,
\\
  T_{+}  &=&  l Q_{+} ,
~~~~
  T_{-} =  -(l + 1) Q_{-},
\\
  Q_{+}  &=&
  \sqrt{\frac{(l+1)^2-m^2}{(2l+1)(2l+3)}} ,
~~~~
  Q_{-}  =
  \sqrt{\frac{l^2-m^2}{(2l-1)(2l+1)}}  .
\end{eqnarray}
We here denote a derivative with respect to  $r$ by a prime. 
The lowest order form of  eq.(\ref{therm0}) is also decoupled 
into the equation of each $l,m $ component, 
\begin{equation}  
 \delta {\dot p} _{l\pm 1 m}  - C^2  \delta {\dot \rho } _{l\pm1 m}
= A C^2 e^\nu \left( e^{ - \lambda } R _{l\pm1 m}
-\frac{3 im \xi _2}{r^2} Q _\pm U_{lm} ^{(1)} 
\right ),
\label{eq.ther}
\end{equation}
with
\begin{eqnarray}
 C^2 & = &  \frac{\Gamma p_0 }{p_0 + \rho _0  } ,
\\
 A & =  & \frac{\rho '  _0
 }{p  _0 + \rho  _0  } - \frac{ p ' _0 }{\Gamma p_0 } .
\end{eqnarray}
In eq.(\ref{eq.ther}), the displacement $ \xi_2 $ 
represents the quadrupole deformation of the stationary star, 
and is related to the quantities of $O(\varepsilon ^2 )$  as
\begin{equation}
  \xi_2 =  -\frac{2}{\nu'}\left(h_2 +
     \frac{1}{3}\varpi^2r^2e^{-\nu}\right) .
\end{equation}
The region for $ A > 0 $ is convectively unstable, while
that for $ A < 0 $ is stably stratified. 
We here assume $ A \ne 0 $ in  the following calculations, but
will separately consider the case $ A =0$ in Section 5.
From  eq.(\ref{eq.ther}), the function $R_{l\pm1 m} $ is 
solved by $ U_{lm} ^{(1)}  $ for the region  $A \ne 0 $.
The function $ V_{l\pm 1 m}  $ describing the horizontal motion
is calculated from the remaining component of eq.(\ref{eq.cons}).
The explicit form is given by 
\begin{eqnarray}
 \left[ l' (l' +1) V_{l' m} \right ]_{ l' =l\pm 1} 
  &=&
  S_{\pm}( v_{2} +  l(l+1) r^2 \Omega e^{-\nu} \dot{U}_{lm} ^{(1)})
  +
  T_{\pm}( v_{1} +  \frac{1}{2} r^2  \varpi e^{-\nu} \dot{U}_{lm} ^{(1)})
\nonumber
\\
 &&~~~~
  +    l(l+1)Q_{\pm} v_{0},
\label{eq.vv}
\end{eqnarray}
where
\begin{eqnarray}
 v_{2}
  &=&
  4 e^{-(3\nu + \lambda)/2}
     \left[  \frac{ r^2 e^{\lambda/2}}{A\nu'}
             \left(
    (e^{\nu/2} \varpi \dot{U}_{lm} ^{(1)} )'
    +
    \frac{\nu'}{2C^2}
     (e^{\nu/2} \varpi \dot{U}_{lm} ^{(1)} )
              \right)
      \right ]'
\nonumber
\\
 &&~~~~
  -  \left(
   \frac{4 r^2 e^{-3 \nu/2}}{\nu'}
   \right)
    (e^{\nu/2} \varpi \dot{U}_{lm} ^{(1)} )'  ,
\\
 v_{1}
  &=&
  2 e^{-(3\nu + \lambda)/2}
     \left[ \frac{ e^{(3\nu + \lambda)/2} }{ A\nu'}
      (r^2 \varpi e^{-\nu})'
      \dot{U}_{lm} ^{(1)}
     \right ]'
  - \left[ 
   \frac{2 e^{-\nu/ 2}}{ \nu'}
   (r^2 \varpi e^{-\nu/ 2})'  
    + \Omega r^2 e^{-\nu} 
     \right ]  \dot{U}_{lm} ^{(1)}  ,
\\
 v_0
  &=&
   \frac{3}{2}\frac{\xi_2}{\varpi}\dot{U}_{lm} ^{(1) \prime} 
   - \left[
    \frac{r^2 e^{-\nu}}{2}( \varpi +2\Omega )
   -   \frac{3}{2}
    \frac{e^{\lambda} m_2}{\varpi r}
   -    \frac{3}{2}e^{-(\nu + \lambda)/2}
    \left[
     e^{(\nu + \lambda)/2} \frac{\xi_2}{\varpi}
     \right]'
   \right]\dot{U}_{lm} ^{(1)}.
\end{eqnarray}
In eq.(\ref{eq.vv}),  we have used eq.(\ref{lap}) to simplify it.

These corrections in the polar functions affect 
the axial parts with indices $(l\pm2, m)$ and $(l, m).$ 
We are interested in the term with index $(l, m)$
as the corrections to the leading equation.
In deriving  the axial equation with these corrections,
the terms up to $O(\varepsilon^3)$ in the background field
also affect it. 
We include both  corrections to the axial equation 
and have the following form,
\begin{equation}
 0  = \dot{U}_{lm} ^{(2)}  - im \chi U_{lm} ^{(2)} 
  + {\cal L}[ \dot{U}_{lm} ^{(1)} ],
\label{eq4.2nd}
\end{equation}
where $ {\cal L} $ is the Sturm-Liouville differential operator
defined by
\begin{eqnarray}
 {\cal L} [\dot{U}_{lm} ^{(1)} ]
  &=&
   8 c_3 \varpi e^{(-\lambda/2 - \nu)}
   (\rho_0 + p_0)
   \left[
    \frac{r^2e^{(\lambda - \nu)/2}}{A\nu'(\rho_0 + p_0)}
    (e^{\nu/2} \varpi \dot{U} _{lm} ^{(1)} )'
    \right]'
  - (F + G) \dot{U}_{lm} ^{(1)} ,
\\
F &=&
    -4 c_3 \varpi^2 e^{(-\lambda/2 - \nu)}
    \left(
     \frac{r^2 e^{\lambda/2} \rho_0 '}{A p_0'}
     \right)' 
  - \frac{4 c_2 \varpi^2 e^{(-\lambda - 3\nu)/2}}{\rho_0 + p_0}
    \left[
    \frac{e^{(\lambda + 3\nu)/2}}{A\nu'\varpi}
    (\varpi r^2 e^{-\nu})'
    (\rho_0 + p_0)
     \right]'
\nonumber 
\\
&&~ 
  +   \left(
    \frac{2 c_1 e^{\nu}}{A\nu'r^2}
    \right)
   [(r^2 \varpi e^{-\nu})']^2  ,
\label{eq4.defff}
\\
G &= &
   -8 c_3 \varpi^2 e^{(-\lambda/2 - \nu)}
    \left(      \frac{r^2 e^{\lambda/2} }{\nu '}
    \right)' 
  + \frac{4 c_2  }{\nu' r^2} (\varpi ^2  r^4 e^{-\nu} )'
\nonumber 
\\
&&~
- 3 c_1
 \left[
  r e^{-\lambda /2 } 
  \left( \frac{ e^{\lambda /2 } \xi_2 }{r} \right) '
  -  \frac{3}{2}\frac{\varpi'}{\varpi}\xi_2
  -  k_2   
 + \frac{e^\lambda}{r} m_2
 + \frac{5W_3}{\varpi}
 \right ]
\nonumber 
\\
&&~
 - \frac{3 m^2}{l(l+1)}
 \left[
  \frac{\xi_2}{r}
  + \frac{1}{2}
   \frac{\varpi'}{\varpi}
    \xi_2
  +    k_2 - \frac{5W_3}{\varpi}
  \right] 
 -\left(   \frac{W_1}{\varpi}  +  \frac{6W_3}{\varpi}  \right ),
\label{eq4.defgg}
\\
 c_{n}  &=&
    \frac{l+1}{l^n} Q^2_{-}
   +(-1)^{n-1}
   \frac{l}{(l+1)^n} Q^2_{+}  .
\end{eqnarray}

In order to solve eq.(\ref{eq4.2nd}),
we introduce a complete set of functions as for the
operator  $ {\cal L } ,$ 
\begin{equation}
   {\cal L } [ y_\kappa   ] + \kappa  y_\kappa  =0  .
\label{eqn4.eig}
\end{equation}
where $ - \kappa $ is the eigenvalue and the eigenfunction 
$ y_\kappa (r) $ is
characterized by $\kappa $, e.g., the number of the nodes.
The eigenvalue is real number of $O(\varepsilon^2)$, 
since  $ {\cal L }$  is the Hermitian operator of $O(\varepsilon^2)$. 
The eigenvalue problem is solved
with appropriate boundary conditions. For
example,  the function should satisfy the regularity
condition  at the center and
the Lagrangian pressure should vanish at the stellar surface.
Certain initial date $f^{(1)} _{lm} $ in eq.(\ref{eq.1st})
can be decomposed by the set. 
We may restrict our consideration to a
single function  labeled by $ \kappa $, 
since the general case is described by discrete sum or
integration over a certain range.
By putting $ y_\kappa = im \chi f^{(1)} _{lm; \kappa}, $
we have 
\begin{equation}
 {\cal L } [ \dot{U}_{lm} ^{(1)}] = 
 {\cal L } [ im \chi f_{lm ; \kappa} ^{(1)} e^{-im(\Omega -\chi)t } ] 
  = -im \kappa \chi f_{lm ; \kappa} ^{(1)} e^{-im(\Omega -\chi)t }  ,  
\end{equation}
where $  ( e^{-im(\Omega -\chi)t } ) ' $ gives the first order
correction and is neglected here.    
Using $ f_{lm ; \kappa} ^{(1)}$,
we can  integrate eq.(\ref{eq4.2nd}) with respect to $ t$, 
and  have the function
$U^{(2)} _{lm} (t,r) $ of $O(\varepsilon^2)$  as
\begin{equation}  
 U_{lm} ^{(2)} = 
 ( i m \kappa  \chi  t f_{lm; \kappa} ^{(1)} +f_{lm} ^{(2)} )
  e^{-im(\Omega -\chi)t } ,
\end{equation}
where the function $ f_{lm} ^{(2)}  $ of  $O(\varepsilon^2)$ 
is unknown at this order.
The sum of the first and second order forms is
approximated as 
\begin{eqnarray}
U_{lm} ^{(1)} + U_{lm} ^{(2)}  &=&
\left[ (1 +  i m  \kappa  \chi t ) f_{lm; \kappa} ^{(1)} 
+f_{lm} ^{(2)} \right] e^{ -im (\Omega - \chi) t} 
\label{eqn4.sum1}
\\
& = &
 \left[ f_{lm; \kappa} ^{(1)} + f_{lm} ^{(2)} \right]
   e^{ -im( \Omega -(1 + \kappa ) \chi ) t } ,
\label{eqn4.sum2}
\end{eqnarray}
where  we have exploited the freedom of $ f_{lm} ^{(2)} $ 
to eliminate the unphysical growing term in eq.(\ref{eqn4.sum1}). 
The value $ \kappa $ originated from
the fixing of the initial data  becomes
evident  for large $t$,
since the accumulation of small effects from the higher order 
terms is no longer neglected.
As a result, the frequency should be adjusted
with the second-order correction
to be a good approximation even for slightly 
large $t$, as in eq.(\ref{eqn4.sum2}). 
This renormalization of the frequency is
closely related to  
treating $t$ as strained coordinate in the perturbation method.
(See e.g. \cite{hi91}.) 
In this way, the specification of the initial data 
at the leading order has influence on the second-order correction
$ \kappa $.

\section{ Oscillations in barotropic stars }

The structure of the neutron stars 
is almost approximated to be barotropic.
The pulsation equation 
is quite different from that of non-barotropic case,
as in the Newtonian pulsation theory. 
The relation (\ref{eq.ther}) for the case $ A=0 $
is replaced by
\begin{equation}
 \delta p _{l\pm1m} = C^2  \delta \rho _{l\pm1 m} =
\frac{p ' _0}{\rho ' _0} \delta  \rho _{l\pm1 m}
= - \frac{ \nu '(p  _0 + \rho  _0) }{2 \rho ' _0} \delta \rho _{lm} .
\label{eq.baro}
\end{equation}
In the last part of eq.(\ref{eq.baro}), we have used
hydrostatic equation of the non-rotating star.
In this case,
the function $R_{l\pm1 m} $ is never determined through 
eq.(\ref{eq.ther}) unlike in $ A \ne 0 $ case, but rather we 
have two restrictions to a single function  $U_{lm} ^{(1)} $.
These conditions are never satisfied simultaneously
unless for $ m= \pm l $, in which one condition is trivial,  
$ \delta p _{l-1 m} = \delta \rho _{l-1 m} =0 $
due to $ Q_{-}=0$. 
The other condition for $ m= \pm l $ becomes
\begin{equation}
 ( e^{\nu /2} \varpi  U_{lm} ^{(1)} ) '
+ \left( \frac{ \nu ' } { 2C^2 } - \frac{ l+1 }{2}
\frac{ ( \varpi r^2 e^{- \nu } ) '}{  \varpi r^2 e^{- \nu } } 
  \right) e^{\nu /2}  \varpi  U_{lm} ^{(1)} 
 = 0 .
\end{equation}
Substituting  the form (\ref{eq.1st}) into this 
and neglecting the higher order term due to  
$ ( e^{ -im (\Omega - \chi) t} )' $, we 
have the same differential equation for $ f_{lm} ^{(1)} .$
The integration with respect to $r$ results in 
\begin{eqnarray}
   U_{lm} ^{(1)} (t,r) &=& f_{lm;\ast} ^{(1)}
 e^{ -im (\Omega - \chi) t} 
\nonumber
\\
    &=&  
 \left[ N_0 ( \rho _0 + p _0 )
 r^{l+1} e^{-\nu} ( \varpi e^{-\nu} ) ^{ (l-1)/2 } 
\right] e^{ -im (\Omega - \chi) t} , 
\label{eq.slu}
\end{eqnarray}
where  $N_0$ is a normalization constant. 
In this way, the function of the lowest order is determined. 
The corresponding the 3-velocity at $ t \to +0$ is given by
\begin{equation}
    \dot {\xi} _{\phi}  = \delta v_{\phi} =
N_0 r^{l+1}  (  \varpi  e^{-\nu} )^{ (l-1)/2 } . 
\label{eq5.velph}
\end{equation}
As shown by \cite{frmo98},
the canonical energy of the perturbation is negative 
for the Lagrange displacement $\xi_{\phi}, $ and the solution 
therefore means 
unstable if the gravitational radiation reaction sets in.

We now  specify  $R _{l\pm1 m}$ of $O(\varepsilon ^2)$ 
to proceed to the pulsation equation with 
the second order corrections.
The function $ X_{l\pm1m} $ describing the radial motion 
is introduced as
\begin{equation}
  R _{l\pm1 m}=  X_{l\pm1m}
+ \frac{3 im e^\lambda  \xi _2}{r^2} Q _\pm U_{lm} ^{(1)} .
\end{equation}
The function $ X_{l\pm1m} $ is arbitrary at this order, but we
have two special cases. 
One is $ X_{l\pm1m}  = 0, $  which is the limiting case 
for eq.(\ref{eq.ther}) in a sense.
The other is chosen  so as to vanish the 
Lagrangian change of the pressure 
within the entire star.
The condition corresponds to   
$X_{l\pm 1 m}  = 4 S _\pm \varpi 
e^{\lambda - \nu} {\dot U }_{lm} ^{(1)} /\nu' .$
For the first choice,  the 
Lagrangian change of the pressure 
never vanishes at the surface unless $ \rho _0 $ vanishes there. 
The second one is the usual way
treated in the Newtonian r-mode,  as shown by  \cite{liip98}.
With the second choice of $ X_{l\pm1m}  $ and eq.(\ref{eq.slu}),  
the pulsation equation  for $ m =\pm l $ leads to 
\begin{equation}
 \dot{U}_{lm} ^{(2)} -im \chi U_{lm} ^{(2)} -G \dot{U}_{lm} ^{(1)}
   =0,
\label{eq5.azero}
\end{equation}
where $G $ is defined in eq.(\ref{eq4.defgg}).
This equation is solved  for  $U_{lm} ^{(2)} $ as in Section 4.
The solution up to $O(\varepsilon ^2) $ is written as
\begin{equation}
  U_{lm} ^{(1)} + U_{lm} ^{(2)}  =
 \left[ f_{lm;\ast} ^{(1)}  + f_{lm} ^{(2)} \right] 
   e^{ -im( \Omega - ( 1 + G) \chi ) t }  ,
\label{eqn5.sum}
\end{equation}
where the function $ f_{lm} ^{(2)} $  of $O(\varepsilon ^2) $ 
is unknown at this order. 
This expression (\ref{eqn5.sum}) is formally the same as 
in eq.(\ref{eqn4.sum2}) for the non-barotropic case
with $\kappa =  G .$ 
The second-order correction in 
the frequency  however depends on the position 
owing to the particular choice of the function 
$  f_{lm;\ast} ^{(1)} .$
%

We show the second-order rotational correction $G$ 
for  $l=m=2$ mode in Fig.1.  We adopt the polytropic 
stellar model with index $ n=1 .$
For the Newtonian star,
$G$ is a positive  function, which monotonically increases 
from the center to the surface.
The value  ranges from $ G = 0.55 (\Omega ^2 R^3/M) $ 
to $ G = 0.75 (\Omega ^2 R^3/M), $
where $ R $ and $ M $ are the radius and the mass for the 
non-rotating star.
These values are rather smaller than that of 
incompressible case, $ G = 37/27 (\Omega ^2 R^3/M).$
(Only for the Newtonian incompressible case, the factor
$ G $ is a constant. See Appendix.)
The stellar deformation $ \xi_2 ,$
which is the most important contribution to $ G ,$ 
diminishes in the compressible fluid.
As the star becomes relativistic,
other relativistic factors become significant. As
a result, 
the factor $ G $ is scaled down as a whole, and eventually
becomes negative for some regions.
In any cases, the frequencies satisfy  the criterion of the 
radiation reaction instability,
which implies  $ 0 < (1 + G) \chi / \Omega < 1 .$
That is, retrograde in the rotating frame and
prograde in the inertial frame. 
Therefore, the second-order correction never
changes  qualitative picture of the instability.

In Fig.2, we show the frequency range of the r-mode oscillations
in the first-order rotational calculation.
The upper and lower limits on the continuous spectrum in  
eq.(\ref{eq.rng}) are shown by two lines. The intermediate values 
between the two lines are allowed for a fixed model $ M/R .$ 
The frequency is a single value $\sigma _N $,  given by 
eq.(\ref{eq.nwfr}) in  the Newtonian limit. 
The  dragging effect relevant to the relativity
broadens the allowed range as shown by eq.(\ref{eq.rng}).
In  Fig. 3, the  allowed range is shown 
including  the second-order corrections
for the extreme  case $ \Omega ^2 = M/R^3 .$
Even in the Newtonian case,
the oscillation frequency is not a single value, since 
the factor $ G $ depends on $ r , $ as seen in Fig.1.
The allowed range of the frequency further increases
with the relativistic factor.
From this  result, 
we expect that the r-mode frequency 
ranges  from $ 0.8 \sigma _N $ to $ 1.2\sigma _N $ 
for a typical neutron star model with
$ M/ R \sim 0.2 .$ 

We examine the effect on the spectrum of the gravitational 
waves emitted by the r-mode oscillations.  
We for simplicity neglect all relativistic corrections
expect in the frequency, and
estimate the spectrum by the Newtonian radiation theory.
The dimensionless gravitational amplitude $h(t)$ at infinity is
determined by evaluating the time variation of the 
current multipole moment,
$ h(t) \sim d^l S_{lm}/dt^l $ (\cite{th80}).
The current multipole moment $ S_{lm} $ for $l=m=2$ mode and
the  velocity (\ref{eq5.velph}) is given by
\begin{equation}
 S_{22} =  N \int \rho _0 (\varpi e^{-\nu} )^{1/2}
e^{-2i( \Omega -\chi)t } r^6  dr ,
\end{equation}
where the normalization $ N$ is also
included the constant from the integration over angular parts.
The Fourier component $ h (\sigma ) $ can be expressed as
\begin{equation}
h( \sigma ) = \int h(t) e^{+i \sigma t } dt 
 \propto  \int  \rho _0 (\varpi e^{-\nu} )^{1/2}
(\Omega -\chi)^2 \delta( \sigma - 2(\Omega -\chi) ) r^{6} dr .
\end{equation}
The spectrum has a finite line breadth as shown  in Fig.4.
The spectrum is non-zero
only for $\sigma  = \sigma _N  $ in the Newtonian treatment, 
but is broad in the relativistic one.
The width at the half maximum is   
$\Delta \sigma / \sigma _N \sim 0.1 .$

\section{Discussion}

In this paper, we have calculated the r-mode oscillations
in the relativistic rotating stars, neglecting 
the gravitational perturbations.
The evolution can be described by oscillatory solutions
which are neutral, i.e., never decay or grow
in the absence of the dissipation.
The oscillation is described not by a single frequency, 
but by the frequencies of a broad range,
unlike in the Newtonian case.
The reason is that the local rotation rate depends on the 
position due to the frame dragging effect even for uniform 
rotation. 
The r-mode oscillation frequency forms a continuous spectrum 
within a certain range. 
This property is distinguished from the well-known
stellar oscillation modes such as the polar f-, p-modes, 
in which  the spectrum of frequency is discrete.
The r-mode frequencies lie in the unstable region for the
gravitational radiation reaction instability, but
it will be an important issue 
whether or not the different spectrum of the unstable modes
leads to different growth, e.g., in the non-linear regime.

\acknowledgments
   We would like to thank Prof. Y. Eriguchi for enlightening 
discussion about this topics.
This was supported in part
by the Grant-in-Aid for Scientific Research Fund of
the Ministry of Education, Science and Culture of Japan
(08640378).

\appendix
\section{Newtonian limit}

In this appendix, we will consider 
the r-mode oscillations in the  Newtonian  limit,
in which 
\begin{equation}
 ( e^{\nu}, e^{\lambda},  (\rho_0 + p_0)/ \rho_0, 
  \varpi  / \Omega )  \to 1, ~~
 (\varpi ',  m_2, k_2, W_1, W_3 )  \to 0. 
\end{equation}
Equation (\ref{eq4.2nd}) reduces to the 
equation derived by \cite{prbero81}, 
if the variables are matched.
They solved the eigenvalue problem,
assuming 
the form $ e^{-i(m \Omega - \sigma_0 (1  + \sigma_1))t }$,
where $\sigma_0$ of $O(\varepsilon)$ is
given by $  2m \Omega /(l(l+1)) $.
The correction $\sigma _1$ of  $O(\varepsilon^2)$
was determined by solving the  eigenvalue problem 
for the operator ${\cal L}_N = D_N - (F_N +G_N)  $ 
in  the  case of  $ A \ne 0 $, 
\begin{equation}
 {\cal L}_N  [y ] + \sigma _1 y  =0,
\end{equation}
with
\begin{eqnarray}
 {\cal D} _N [y ]
  &=&
   4 c_3 \Omega ^2  \rho_0  \left[
    \frac{r^2 }{A g \rho_0 } y'  \right ]',
\\
  F_N &=&  
  -4 c_3 \Omega^2  \left( \frac{r^2 \rho_0 '}{A p_0 '} \right)'
  - \frac{4 c_2 \Omega^2 }{\rho_0 }
  \left[ \frac{\rho_0 r }{A g }  \right]'  +
  \left( \frac{4 c_1 \Omega^2 }{A g } \right),
\\
G_N &= &
   -4 c_3 \Omega^2  \left( \frac{r^2 }{g} \right)'
  + \frac{8 c_2  \Omega ^2  r}{g} 
  + 2 c_1  r  \alpha  '  + \frac{2 m^2}{l(l+1)}  \alpha ,
\end{eqnarray}
where we have used the gravitational acceleration 
$  g = \nu ' /2 $ and
ellipticity $ \alpha = - 3 \xi_2/( 2r) .$
The second-order correction $ \sigma _1$
exactly corresponds to $ \kappa $ in eq.(\ref{eqn4.eig}).

%
As shown previously (\cite{prbero81},  \cite{sa82}),
the eigen-value problem becomes singular\footnote{
Recently, \cite{lofr98} showed 
the resolution of the singularity in Newtonian 
isentropic stars.}
for  the barotropic case $ A =0 $,
since the second-order differential term vanishes.
The second-order solution (\ref{eqn5.sum}) 
in the Newtonian limit reduces to 
\begin{equation}
 U_{lm} ^{(1)} + U_{lm} ^{(2)}  =
 \left[ N_0 \rho _0 r^l   + f_{lm} ^{(2)} \right]
   e^{-i(m \Omega - \sigma_0 (1  + \sigma_1))t }.
\end{equation}
This equation is valid only for $l = \pm m$, since the Newtonian
counterpart of eq.(\ref{eq.baro}) is never satisfied otherwise.
When the correction $ G_N $ is not a constant,
the eigenvalue  problem is ill-posed.
The function $ G_N $ indeed depends on  $r$ for the 
compressible matter, 
so that \cite{prbero81} 
concluded no solution in this case.
The exceptional case is the incompressible fluid,
in which $ G_N $ is a constant since 
$ \alpha = 5/4 (\Omega^2 R^3/M), g = M r/R^3 .$
The correction in the frequency for $ l =m $ is 
\begin{equation}
 G_N =  - \frac{4l}{ (l+1)^3 } \frac{ \Omega^2 R^3}{M} 
+ \frac{2 l }{  l+1 }  \alpha ,
\end{equation}
which should be the same value calculated 
by \cite{prbero81}
except a misprint in their expression.



\clearpage

\begin{figure}[h]
 \plotone{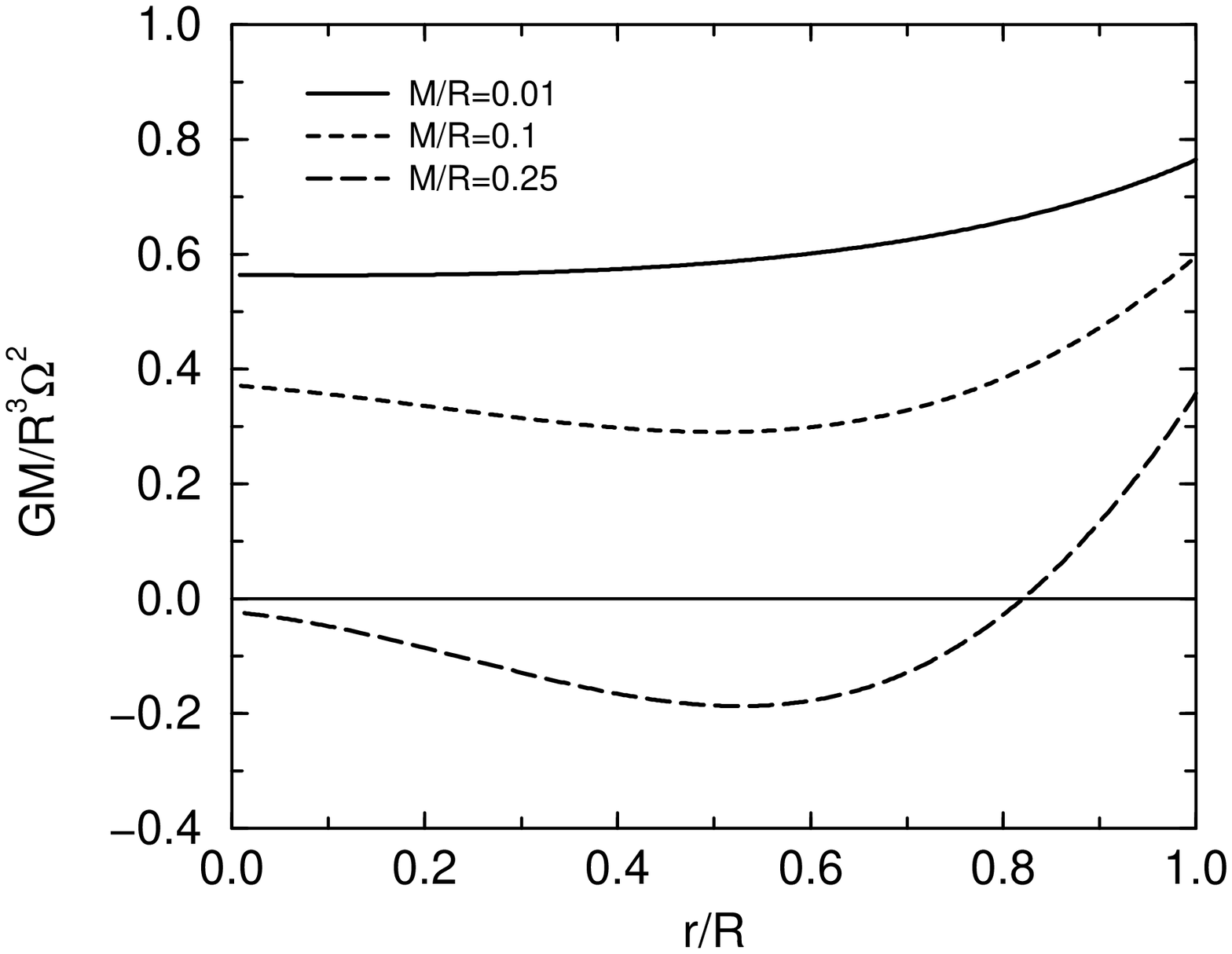}
\label{fig1}
\figcaption[f1.eps]{
The second-order correction in the frequency as a function of radius.
The factor $ G $ is normalized by using the mass $ M $, radius $ R$ 
and angular velocity $ \Omega  .$ 
The solid line is for the Newtonian case $ M/R =0.01 .$
The dotted line represents the result for $ M/R =0.1,$
and the dashed line for  $ M/R =0.25.$
}
\end{figure}
\clearpage

\begin{figure}[h]
 \plotone{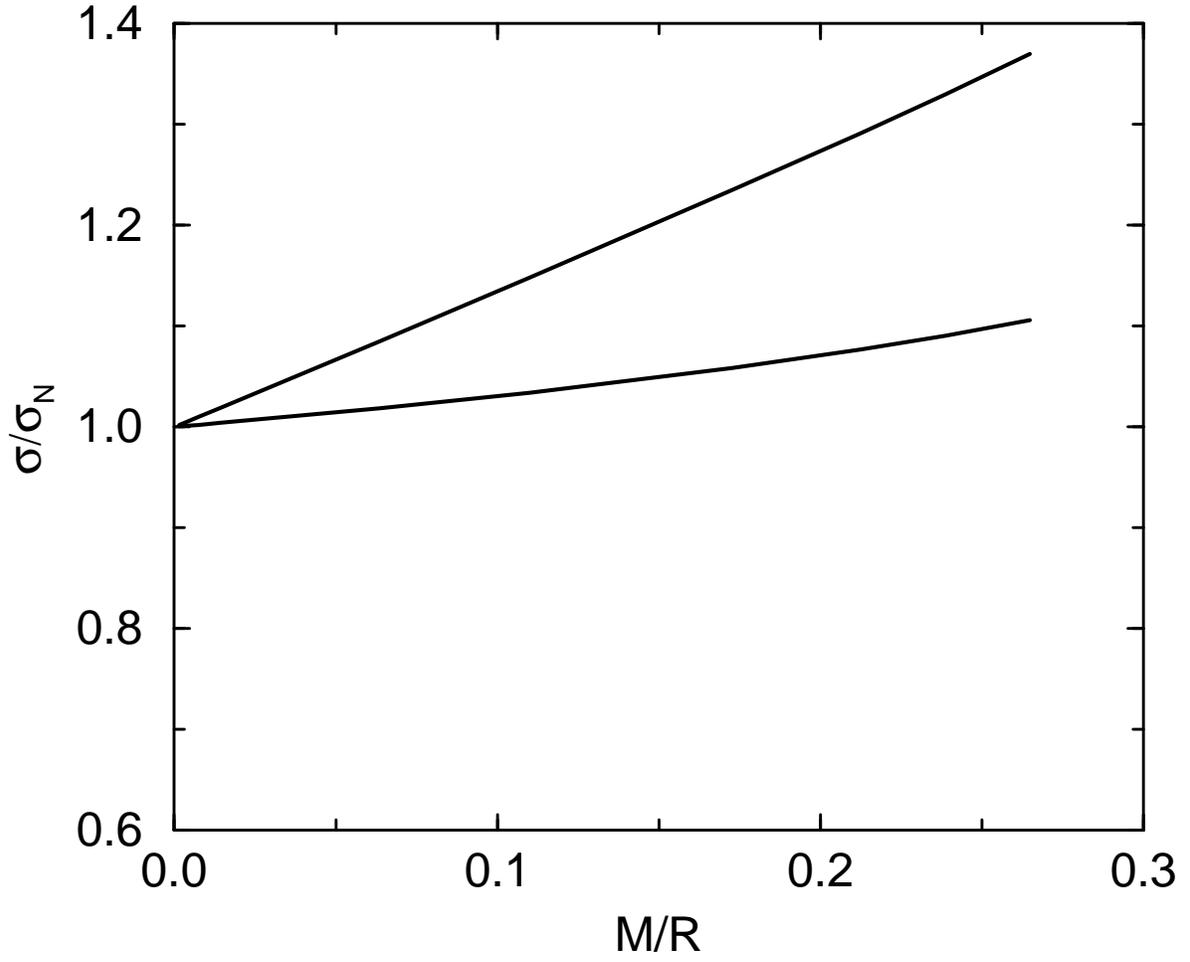}
\label{fig2}
\figcaption[f2.eps]{
The frequency range  with the relativistic factor $ M/R .$
Two lines denote the upper and lower limits of the frequency.  
The intermediate values between them are allowed
for a fixed model $ M/R .$ 
The frequency is normalized by the Newtonian value $ \sigma _N .$
}
\end{figure}
\clearpage

\begin{figure}[h]
 \plotone{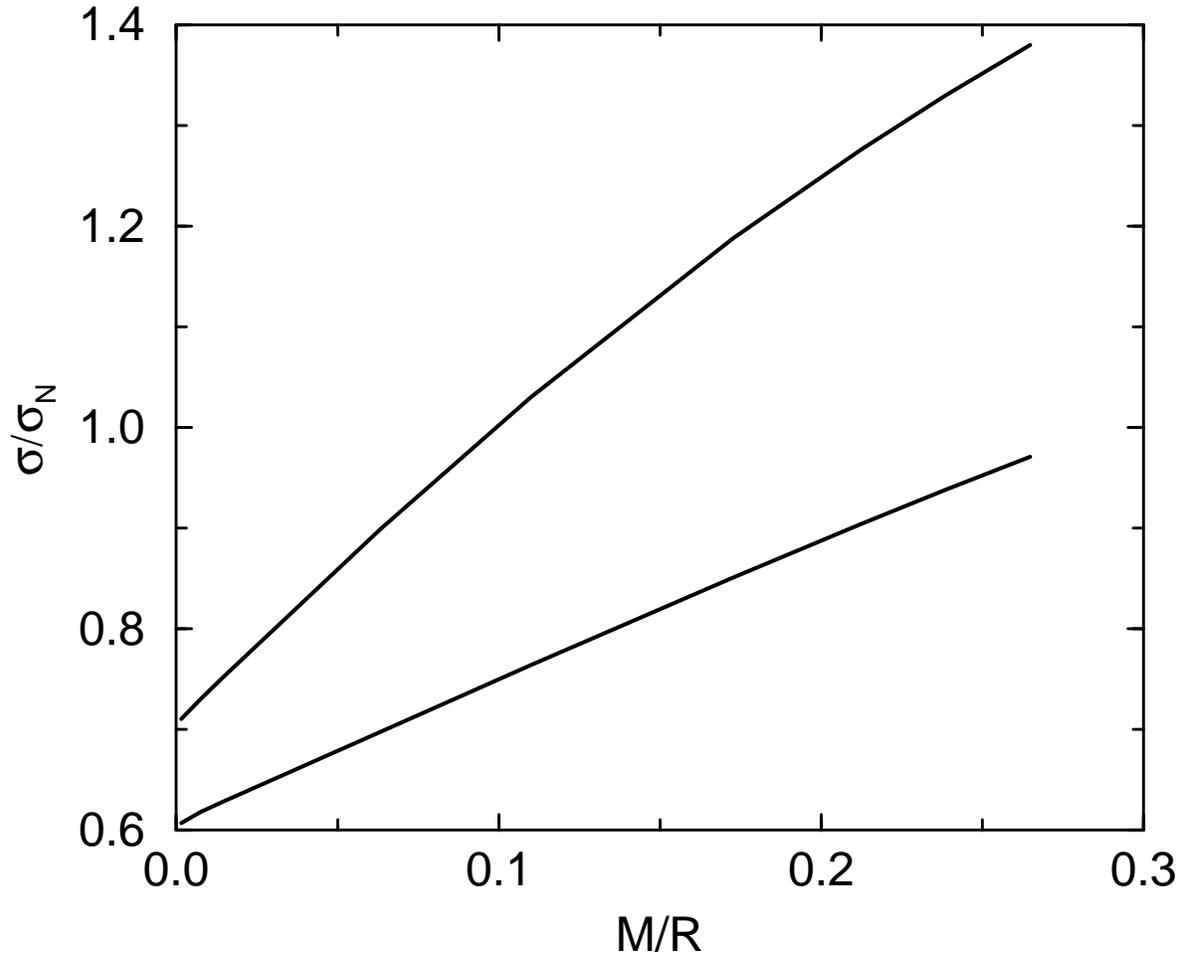}
\label{fig3}
\figcaption[f3.eps]{
The same as Fig.2, but 
the second-order correction with $\Omega ^2 = M/R^3$
is included.
}
\end{figure}
\clearpage

\begin{figure}[h]
 \plotone{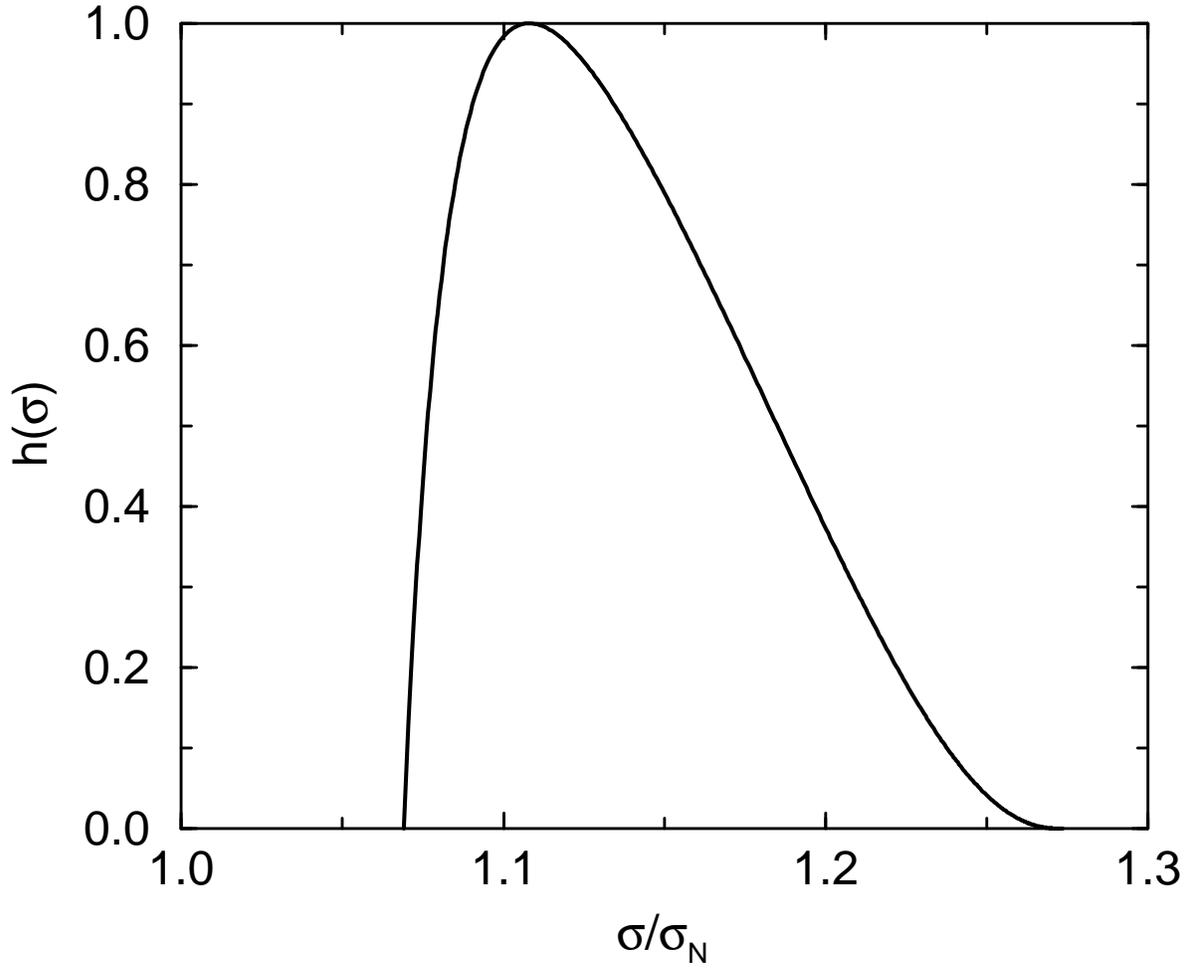}
\label{fig4}
\figcaption[f4.eps]{
The Fourier transform of the gravitational wave amplitude
of $ l=m=2$ mode for stellar model $ M/R =0.2 .$
The frequency is normalized by the Newtonian value $ \sigma _N ,$
and the amplitude $h(\sigma)$ is normalized by the maximum.
}
\end{figure}

\end{document}